\documentclass[aps,epsf,floats,epsfig, twocolumn,byrevtex, showpacs]{revtex4}
\usepackage{amsfonts}
\usepackage{amsmath}
\usepackage{amssymb}
\usepackage{graphicx}

\newcommand{\beq}{\begin{equation}}
\newcommand{\eeq}{\end{equation}}
\newcommand{\beqn}{\begin{eqnarray}}
\newcommand{\eeqn}{\end{eqnarray}}
\newcommand{\bearr}{\begin{array}}
\newcommand{\enarr}{\end{array}}

\begin{document}
\newcommand {\ee}[1] {\label{#1} \end{equation}}
\newcommand{\be}{\begin{equation}}

\preprint{ }

\title{From multiplicative noise to directed percolation in wetting
transitions}
\author{ F. Ginelli$^{1,2}$,V. Ahlers$^{3}$, R. Livi$^{4,2}$,
D. Mukamel$^{5}$, A.~Pikovsky$^{3}$, A. Politi$^{1,2}$ and A. Torcini$^{1,2}$}
\affiliation{
$^1$Istituto Nazionale di Ottica Applicata,
           Largo E. Fermi 6, Firenze, I-50125 Italy \\
$^2$ Istituto Nazionale di Fisica della Materia, Unit\`a di Firenze, Italy\\
$^3$ Department of Physics, University of Potsdam, Potsdam, Germany \\
$^4$ Dipartimento di Fisica, Universit\`a di Firenze, Italy \\
$^5$ Department of Physics of Complex Systems, The Weizmann
Institute, Rehovot, Israel}
\date{\today}
\widetext
\begin{abstract}
A simple one-dimensional microscopic model of the depinning transition of an
interface from an attractive hard wall is introduced and investigated. Upon
varying a control parameter, the critical behaviour observed along the
transition line changes from a directed-percolation to a multiplicative-noise
type. Numerical simulations allow for a quantitative study of the multicritical
point separating the two regions, Mean--field arguments and the mapping on a
yet simpler model provide some further insight on the overall scenario.
\end{abstract}


\maketitle

A variety of interesting physical phenomena corresponds to the
unbinding transition of an interface from a flat surface. This is
the case of wetting processes (WP) taking place in the thin liquid
film which forms on a substrate exposed to a gas. By varying
external parameters such as temperature or pressure, the liquid
layer $h(x,t)$ may either grow and become macroscopically thick or remain
confined to the close vicinity of the substrate \cite{Diet,HH}.
Wetting phenomena can also take place under non-equilibrium
conditions. Here one is interested, for example, in a growth
process of a film over a substrate. Depending on the dynamical
rates controlling the growth process one can observe similar
pinned or unpinned phases.

A question of general interest concerns the universality of the
unbinding transition. While the equilibrium scenario is well
established \cite{Diet}, an overall understanding of
non-equilibrium wetting phenomena is still lacking. Numerical
studies of non-equilibrium systems have revealed a composite
picture that still needs to be fully disentangled. Here one
considers the dynamical equations of a moving interface
interacting with a hard wall. In the simplest case, the unbinding
transition is signalled by the change of sign of the average
velocity of the free interface. A pinned phase is obtained when
the free interface moves towards the substrate and an unpinned
phase is found when it moves away. This scenario is sometimes
referred to as the multiplicative noise (MN) and is well described
by a Kardar-Parisi-Zhang (KPZ) equation ~\cite{KPZ} with a
hard-wall \cite{MN,MN2,MuHwa,Munoz}. On the other hand, a study of
some discrete growth models has shown that for particular values
of the growth rates the unbinding transition is of different
nature, belonging to the directed percolation universality class
\cite{Alon}.

More interesting is the scenario when the surface exerts, in
addition, a short range attractive force that may prevent $h$ from
growing even when the "free" interface would otherwise have a
positive velocity. In WP such an effective interaction is obtained
when the growth rate on the bare substrate is lower than the
growth rate on the film itself. Some microscopic models,
introduced to study this out--of--equilibrium depinning transition
\cite{SOS1,SOS2}, indicate that for a sufficiently strong
attractive force, the unbinding transition may become
discontinuous.

This scenario is by no means restricted to WP. Similar features
have been, e.g., observed in the onset of complete synchronization
in chains of coupled oscillators\cite{Baroni,Ahlers,Ginelli}. In
this context, two replicas of the chain are either subject to the same
stochastic forcing, or locally coupled: the absolute difference
$\Delta(i,t)$ between the state variables in the $i$th site
may either go exponentially to zero, in which case
synchronization eventually occurs, or stay finite. Upon formally
introducing $\gamma(i,t) = - \log \Delta(i,t)$, it is easily
recognized that $\gamma(i,t)$ plays the same role as $h(x,t)$:
synchronization corresponds to the unpinned phase, while the
unsynchronized regime corresponds to the bounded phase. In this
context, a strong nonlinearity may play the role of the attractive
force in WP, preventing small but finite perturbations from
vanishing. However, at variance with WP, in complete
synchronization, the effect of a sufficiently strong ``attractive
force'' is to bring the MN transition into the directed
percolation (DP) universality class \cite{Baroni,Ahlers} rather
than making it first order.

In the absence of a sufficiently general field-theoretic approach
able to reconcile all the various observed scenarios into a common
framework, the study of minimal models is very helpful for the
identification of the basic mechanisms. This is the main
motivation for introducing hereafter a simple microscopic model.
By numerically reconstructing its phase diagram, we shall be able
to clearly recognize that both the MN and DP universality classes
can be found and to quantitatively investigate the
``multicritical'' point separating the two scenarios. Moreover, an
accurate reconstruction of the critical line will be proposed
based part on mean-field arguments and exploiting the exact
mapping (in a limit case) onto a genuine DP-model.

Specifically, we consider a simple growth model, which, for
reasons that will become clear in a while, is called {\it
Single--Step--plus--Wall} (SSW) model. The starting point is the
Single--Step (SS) model originally introduced to study the
roughening of 1d interfaces \cite{MRSB,KS}. It is well known that
the SS model can be exactly mapped onto the 1d KPZ
equation\cite{MRSB}. The interface is described by a set of
integer heights $h_i$ on the sites $i$ of a one-dimensional
lattice of length $L$, satisfying the ``continuity'' constraint
$|h_i-h_{i+1}|=1$. At each time step, $dt=1/L$ a site $i$ is
randomly selected and its height increased, $h_i \to h_i +2$,
provided that a local minimum exists at site $i$. In the
thermodynamic limit, since the dynamics does not introduce any
spatial correlation, a generic interface with mean 
slope $s$ moves with a mean
velocity $v(s) = (1-s^2)/2$. The exact knowledge of $v(s)$ will be
crucial in the following, since it allows for an exact
determination of the critical line in the MN regime.

The second ingredient of the SSW model is an upward--moving wall located at
some integer height $h_w(t)$ below the SS interface. It moves with velocity
$v_w$ and both ``pushes'' and attracts the interface. Altogether, the SSW
dynamics amounts to the following evolution algorithm: at each time step, a
site $i$ is randomly chosen and, if it is a local minimum, $h_i$ is
increased by two units with probability $1$, or $(1-q)$, depending
whether $h_i > h_w$, or $h_i = h_w$ (see Fig.~\ref{SS2}a). After $n_w=L/v_w$
steps, the wall is moved upwards by one unit and, simultaneously, the height
of the interfacial sites overtaken by the wall, is increased by two units
(see Fig.~\ref{SS2}b).
\begin{figure}[tcb]
\centering
\includegraphics*[width=6cm, angle=0]{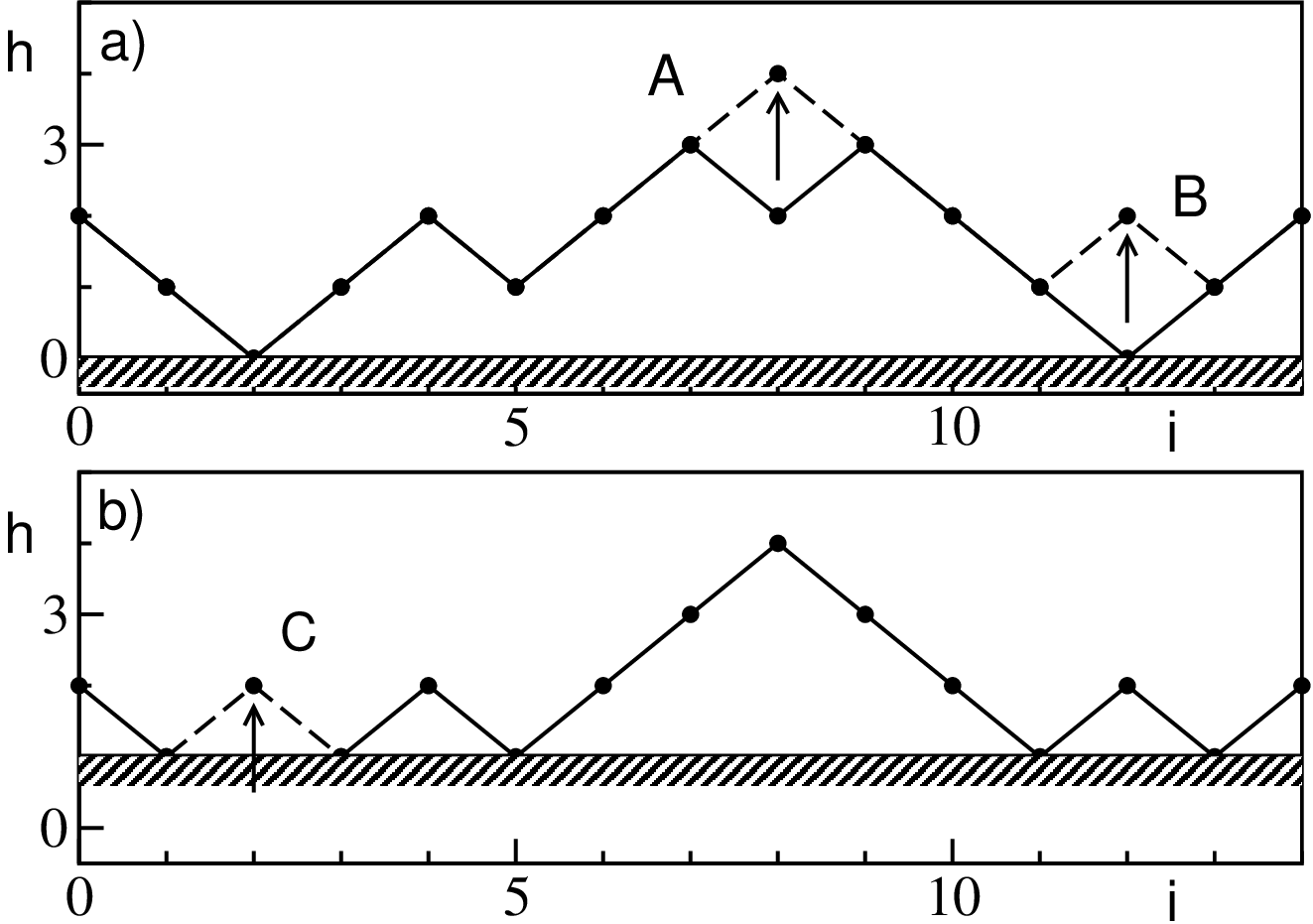}
\caption{Updating rule of the SSW model. The full line represents the interface,
while the shaded area identifies the wall. Dashed segments indicate interface
flips occurring in randomly chosen local minima (see A and B in panel a) and in
all sites located below the wall after it has been shifted upwards by one unit
(see C in panel b).}
\label{SS2}
\end{figure}
Physically, the SSW describes a roughening and moving interface,
attracted by a short-range force to a hard wall. Its dynamics is
determined by two parameters: (i) the relative velocity of the
wall with respect to the free-interface, that we control by
modifying $v_w$,\cite{note}; and (ii) the stickiness of the wall,
quantified by $q$. Since we are interested in characterizing the
phase diagram of SSW by locating the depinning transition from the
wall, the natural order parameter is the density of sites pinned
at the wall
\beq \rho(t) = \frac{2}{L} \left\langle \sum_{i=1}^{L} s_i(t)
\right\rangle
,
 \quad s_i(t) = \
 \left\{\begin{array}{ll}
1 & {\rm if} \enskip h_i(t) = h_w\\
0 & {\rm if} \enskip h_i(t) > h_w
\end{array}\right.
\label{rho}
\eeq
where $\langle\cdot \rangle$ denotes an ensemble
average over different realizations of the stochastic process.

A necessary condition for interface depinning to occur is
$v_w<1/2$, because $1/2$ is the velocity of a free and flat interface. For $q$
small enough, this is also a sufficient condition, since the
attractive force is overcompensated by the faster velocity of an almost
straight interface. This can be seen by a simple mean field argument:
an interface completely attached to the wall has a higher density
of minima (1/2) than a rough one (1/4), so that its average
velocity $v = (1 - q)$ is larger than 1/2 as long as $q < q^* =
1/2$. Accordingly, below $q^*$, a pinned interface detaches as
soon as $v_w<1/2$. Numerical simulations confirm that the
transition indeed occurs at $v_w=1/2$, with the only slight
difference that $q^* = 0.4445(5)$. Above $q^*$ the interface may
remain pinned even when its velocity is larger than $v_w$ (see
Fig.~\ref{PhasePlane}). In other words there is a sort of bistable
region, where an initially pinned interface remains attached while
a depinned one moves away from the wall. The transition line,
located at the lower border of the bistable region, is continuous
and turns out to belong to the DP universality class. This is at
variance with the discontinuous transition observed, e.g., in the
solid-on-solid model of \cite{SOS2}. Before commenting on the
possible reason of such a difference, it is necessary to explore
in more quantitative way the critical behavior both above and
below $q^*$. 

Continuous non-equilibrium phase transitions are characterized by three
independent critical exponents. At criticality ($v_w = v_w^c$), the density
$\rho(t)$ of pinned sites scales with time as $\rho \sim t^{-\delta}$, while
its stationary value depends on the the distance from criticality as,
\begin{equation}
\lim_{t \to \infty} \rho(t) \sim (v_w - v_w^c)^{\beta}, 
\label{power1}
\end{equation}
Very accurate numerical estimates of the DP critical exponents
give $\beta_{DP} = 0.276486 \pm 6 \cdot 10^{-6}$ and $\delta_{DP}=
0.159464 \pm 6 \cdot 10^{-6}$  \cite{Jensen}. Less accurate
estimates are available for the MN scenario, namely $\beta_{MN}=
1.7 \pm 0.1$ and $\delta_{MN} = 1.1 \pm 0.1$ \cite{MN2}. The third
exponent, $z$, can be defined with reference to the scaling
relation $\rho(t,L) = L^{-\delta z} g\left(t\, L^{-z}\right)$
\cite{HH}, where $g$ is a proper scaling function. It takes very
similar values in DP and MN, namely, $z_{DP}=1.580745 \pm 10^{-6}$
while $z_{MN}=1.53 \pm 0.07$.

We have performed numerical simulations of the SSW model for different
$q$ values starting from an initially pinned ($\rho (0) = 1$) interface.
The exponents $\beta$ and $\delta$ have been measured by studying large system
sizes (from $L=2^{17}$ to $L=2^{20}$) in order to minimize finite-size
corrections, and by averaging over a small number of different realizations
($\approx 10$) to further decrease statistical fluctuations. Conversely,
it is sufficient to consider much smaller sizes ($2^5 < L < 2^{10}$) for
a reliable estimate of $z$ that has been determined by looking for the optimal
collapse of the various $\rho(t,L)$ curves (in this case, however, it has been
necessary to average over a much larger ensemble of realizations ($>10^4-10^5$).

A complete summary of the resulting values of the critical exponents is
reported in Table~\ref{SSWtable}. Altogether, we find that for $q < q^*$, the
attracting force not only does leave the transition point unaffected, 
but also
the universality class of the critical behaviour remains of MN type
\cite{foot1}. For $q>q^*$, the transition line veers down, while the critical
exponents signal a transition of DP type.

\begin{figure}[tcb]
\centering
\includegraphics*[width=6cm, angle=0]{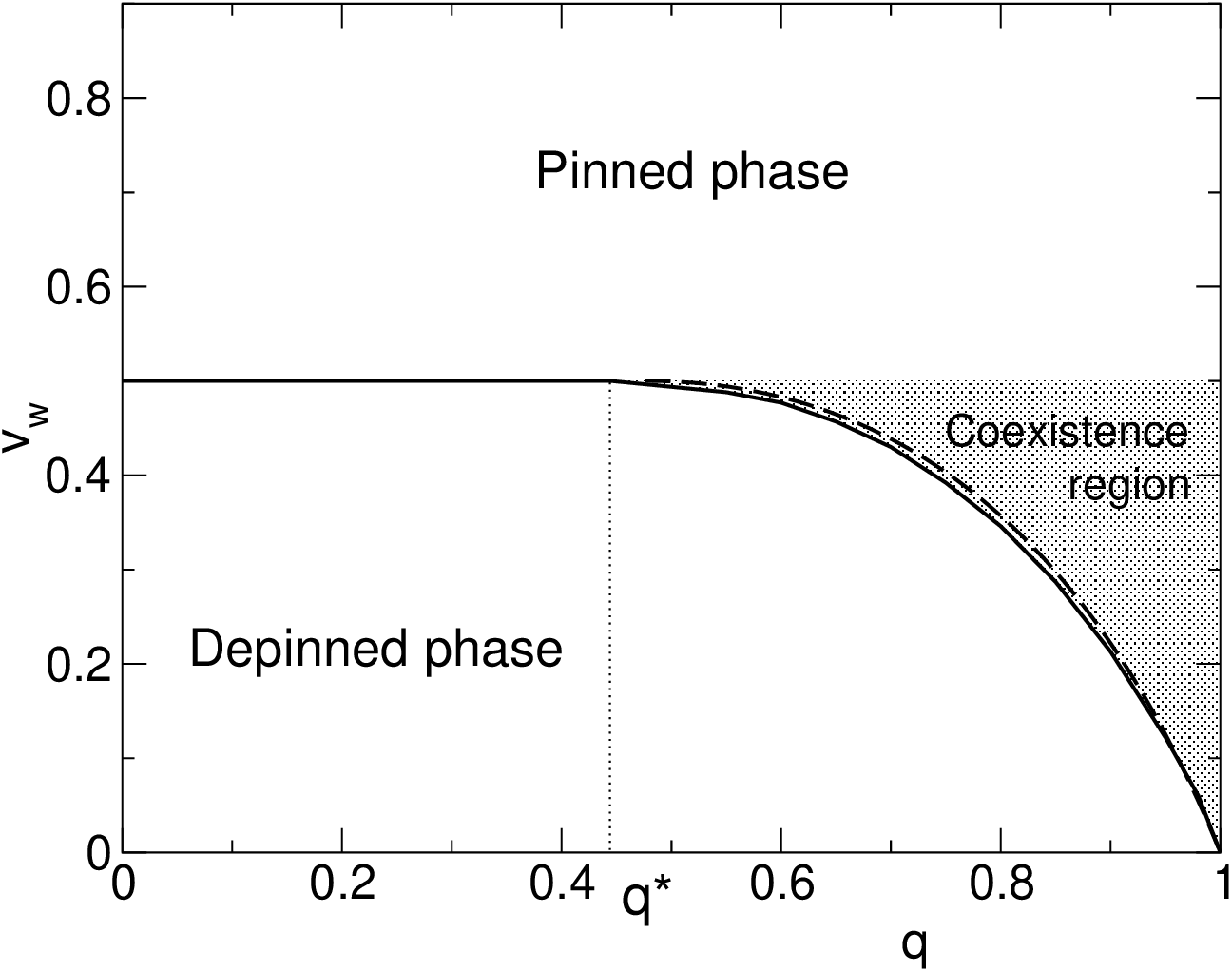}
\caption{Phase diagram of the SSW model. The depinning transition takes
place along the solid line; the dashed line is the result of the approximate
analytic mapping discussed in the text. In the shaded area, stationary pinned
interfaces exist even though $v_w$ is smaller than the free interface velocity
(=1/2).}
\label{PhasePlane}
\end{figure}

\begin{table}[tbh]
\centering
\begin{tabular}{||c|c|c|c|c||} \hline
$q$ & $v_w^c$ & $\delta$ & $\beta$ & $z$ \\ \hline
0    &  0.5  & $1.14(5)$   & $1.70(5)$ & $1.5(1)$  \\ \hline
0.2  &  0.5  & $1.13(5)$   & $1.75(5)$ & $1.5(1)$  \\ \hline
0.4  &  0.5  & $1.12(5)$   & $1.5(1)$  & $1.3(1)$   \\ \hline
$0.4445(5)$  &  0.5     & $0.50(1)$  & $0.74(5)$ & $1.5(1)$ \\ \hline
0.6  & $0.47635(5)$   & $0.15(1)$ &  $0.276(5)$  & $1.5(1)$ \\ \hline
0.7  &  $0.42975(5)$  &  $0.16(1) $  &  $0.27(1) $ & $1.5(1)$  \\ \hline
0.8  & $0.34895(5)$   & $0.17(1)$   & $0.276(5)$   & $1.5(1)$ \\ \hline
\end{tabular}
\caption{Critical exponents of the SSW model. In parentheses we report the
estimated uncertainty on the last figure.}
\label{SSWtable}
\end{table}

The DP critical line can be best understood by analyzing the SSW
model in the vicinity of the point $q=1$ and $v_w=0$. Here, the
dynamics is dominated by two slow mechanisms: (i) detachment of
pinned sites during the asynchronous part of the rule; (ii)
shrinking of the unpinned islands at the wall move. In comparison,
the dynamics of detached regions between consecutive wall moves
rapidly leads them to assume a perfectly triangular shape with a
maximal slope equal to $\pm 1$. Therefore, such islands correspond
to the dead phase in DP, since their shape prevents the occurrence
of any pinning in their interior. It is now convenient to divide
between attached ($h_i=h_w$) and detached ($h_i > h_w$) sites,
denoting them with $A$ and $D$, respectively. Correspondingly, the
dynamics reduces to a simple probabilistic cellular automaton:
${L}/{v_w}$ sites are first randomly selected, transforming each
$A$ into a $D$ with probability $1-q$; next, the wall move amounts
to transforming all $A$'s into $D$'s and all $D$'s neighbouring an
$A$ into a $D$ (see Fig. \ref{SSW3}). Apart from the peculiar presence
of both an asynchronous and a synchronous part, this rule clearly
belongs to the class of contact processes with an absorbing state
($D$) and, as such, it is expected to exhibit a DP
transition\cite{Gras,janssen}. The only relevant parameter is the
ratio between the rate $(1-q)$ and the wall velocity $v_w$, which
corresponds to the slope of the critical line at $q=1$. In fact,
numerical simulations of the automaton yield a ``critical ratio''
$a_c =\lim_{q\to 0} {v^c_w}/({1-q}) = 2.866\ldots$, in very good
agreement with the slope determined from direct simulations of the
SSW model.

\begin{figure}[tcb]
\centering
\includegraphics*[width=6cm, angle=0]{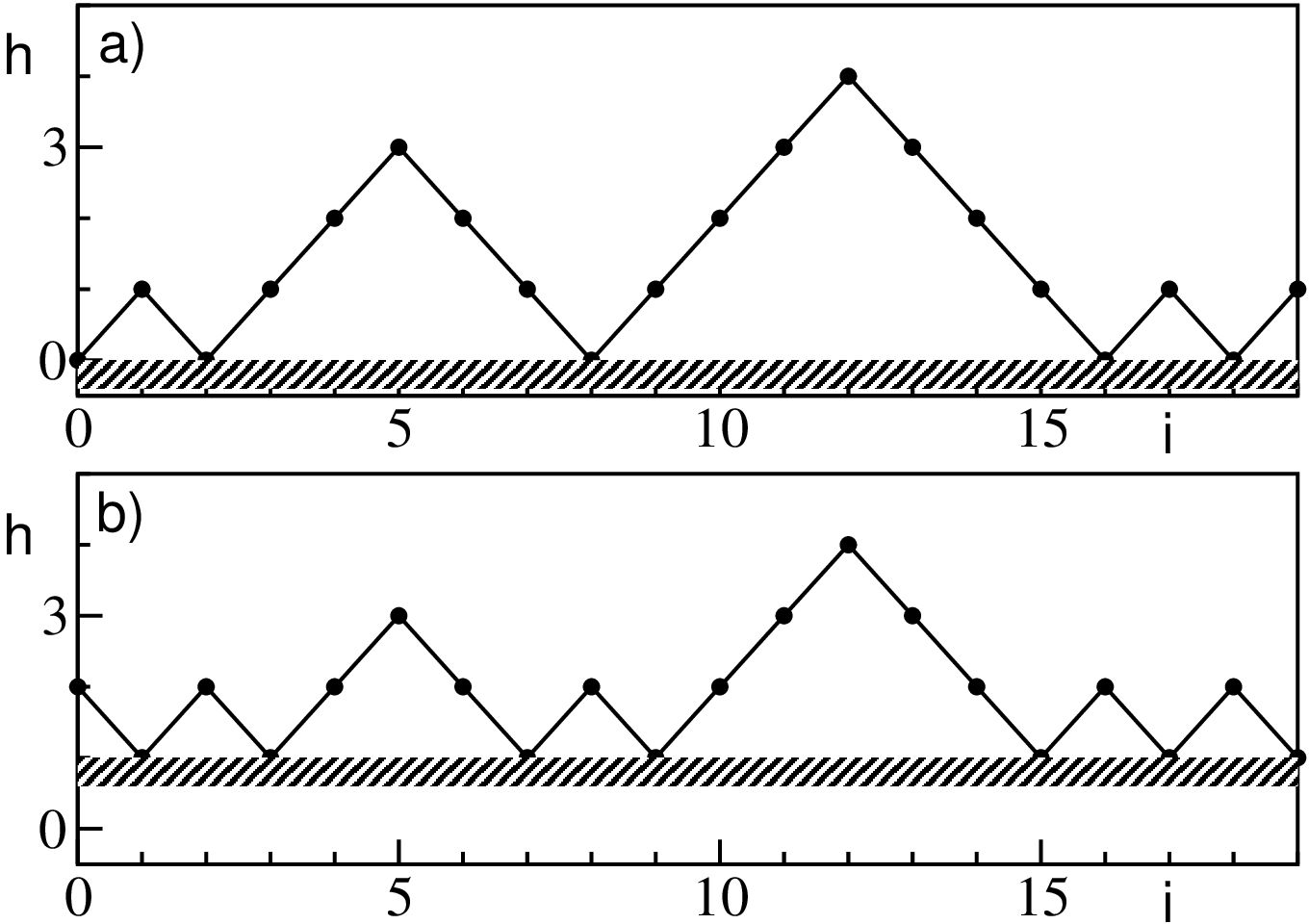}
\caption{
The interface before (panel a) and after (b) the wall move, with the same
settings as in Fig.~\ref{SS2}. The corresponding configurations read 
as $ADADDDDDADDDDDDDADAD$ and $DADADDDADADDDDDADADA$ (see text for the
definition of $A$ and $D$.)}
\label{SSW3}
\end{figure}

At a finite distance from $q=1$, the automaton does no longer
describe exactly the SSW dynamics, since fluctuations of the
interface within an unpinned island can both induce a faster
shrinking of the island and the generation of pinned sites in its
interior. The first effect amounts to increasing by some factor
$\alpha$ the average number of $D$'s turned into $A$'s at the island
borders, so that the critical point would be determined by the
equation
\beq \frac{\alpha v^c_w}{1-q} = a_c \quad .
\label{alp}
\eeq
In the configuration plotted in Fig.~1, the island border
identified by the letter $C$ shifts by three units thus implying
$\alpha =3$. The average value of $\alpha$ can be determined by
noticing that, at criticality, in the SSW model, the typical slope
$s$ of the interface inside an island must be such that its
velocity coincides with the wall velocity $v^c_w$. If we then
assume that the profile is a biased random walk with probabilities
$p_u$ and $1-p_u$ of up and down moves, respectively, one finds,
using simple combinatorial considerations, $\alpha =
1+2(1-p_u)/p_u$. Since $p_u$ is simply related to the slope, $s$,
by $p_u = (s+1)/2$ one finally obtains $\alpha = 1 +
2(1-s)/(1+s)$. Inserting this expression into (\ref{alp}) and
eliminating $s$ with the help of the relation $v^c_w = (1-s^2)/2$,
one obtains $v^c_w(q)= 2 \sqrt{1 + 2 a_c (1-q)} - a_c (1-q) - 2$.
Although approximate, this formula reproduces very accurately the
DP critical line not only in the vicinity of $q-1$ but also up to
the multicritical point, where it touches the MN critical line
(see Fig.~\ref{PhasePlane}), providing a good approximation for
its position as well ($q_{\rm max} = 1 - 3/(2a_c) = 0.477\ldots$).
Evidently, the quality of the theoretical formula implies that
even close to the multicritical point, the sudden appearance of
pinned sites inside unpinned islands does not significantly modify
the transition point. A deeper understanding of this point is left
to future investigations.

The correspondence with MN and DP critical phenomena unveiled for
small and large $q$ values, respectively, is not sufficient to
make predictions about the scaling behaviour in the vicinity of
the multicritical point\cite{foot2}. There, in the absence of a
convincing field-theoretic approach, a chance for understanding
how the two out-of-equilibrium critical phenomena may be connected
to one another is offered by numerical investigation. However,
even this is not a straightforward task, since three levels of
criticality mix together: criticality of the free rough interface,
criticality of the depinning transition and, finally, that one
connected with the MN-DP transition. While approaching $q^*$ from
the left along the critical line $v_w^c = 1/2$, the power--law
decay of the density of pinned sites $\rho(t)$ turns out to be
first governed by the exponent $\delta^* \approx 1/2$, which
crosses over to $\delta_{MN}$. As the crossover time appears to
diverge when $q \to q^*$, it can be safely stated that $\delta^*$
characterizes the critical behaviour at $q^*$. The same scaling is
expected for $q$ approaching $q^*$ from the right along the
critical line, with a crossover from $\delta^*$ to $\delta_{DP}$.
However, the difficulty of locating the critical line with a
sufficient accuracy prevents an effective numerical verification.
The other two critical exponents are $\beta^* \approx 3/4$ and
$z^* \approx 3/2$ at $q^*$, so that both the $\beta^*$ and the
$\delta^*$ values are intermediate between the corresponding $MN$
and $DP$ critical exponents, while $z^*$ is compatible with both
$z_{MN}$ and $z_{DP}$.

In conclusion, with reference to a simple microscopic model, we have shown that
MN and DP can be different facets of the same wetting process. The connection
between these two different universality classes strongly hints at the
possibility that both out-of-equilibrium transitions may be described within
a single field--theoretic approach. Some progress has been recently made 
in this direction in \cite{Munoz2}, where the authors have found DP behavior 
in a KPZ equation with attractive wall.

\begin{acknowledgments}
We are particularly to P. Grassberger for his
contributions in the early stages of this work. A. Giacometti and
L. Giada are acknowledged for useful discussions. CINECA in
Bologna provided us access to the parallel Cray T3 computer under
the INFM-grant ``Iniziativa Calcolo Parallelo''. This work has
been partially funded by the FIRB-contract n. RBNE01CW3M\_001 and
the Israeli Science Foundation.
\end{acknowledgments}


\begin{thebibliography}{999}
\bibitem{note} The reason for this choice is purely technical: the nonlinear
term of the KPZ equation describing the evolution of the free interface is
proportional to its velocity \protect\cite{MRSB}. Since this term is also
responsible for the out-of-equilibirum dynamics, we prefer to keep it constant.
\bibitem{foot1} For $q=0.4$, the estimate of $\beta$ is not in good agreement
with the known MN value. This can be probably ascribed to finite-size effects
due to the proximity of the multicritical point $q^*$.
\end{thebibliography}
\end{document}